# A Statistical Model for Melody Reduction

Tianxue Hu[1†]

Claire Arthur[2]

[1, 2] Computational & Cognitive Musicology Lab, Georgia Tech Center for Music Technology,
Georgia Institute of Technology, Atlanta, Georgia, United States

[†] Corresponding author: thu82@gatech.edu

## Abstract

A commonly-cited reason for the poor performance of automatic chord estimation (ACE) systems within music information retrieval (MIR) is that non-chord tones (i.e., notes outside the supporting harmony) contribute to error during the labeling process (e.g., Chen & Su, 2018; Masada & Bunescu, 2017). Despite the prevalence of machine-learning approaches in MIR, there are cases where alternative approaches provide a simpler alternative while allowing for insights into musicological practices.

In this project, we present a statistical model for predicting chord tones based on music theory rules. Our model is currently focused on predicting chord tones in classical music, since composition in this style is highly constrained, theoretically making the placement of chord tones highly predictable. Indeed, music theorists have labeling systems for every variety of non-chord tone, primarily classified by the note's metric position and intervals of approach and departure. Using metric position, duration, and melodic intervals as predictors, we build a statistical model for predicting chord tones using the TAVERN dataset (Devaney et al. 2015). While our probabilistic approach is similar to other efforts in the domain of automatic harmonic analysis (e.g., Temperley 1997; Temperley & Sleator 1999), our focus is on melodic reduction rather than predicting harmony. However, we hope to pursue applications for ACE in the future. Finally, we implement our melody-reduction model using an existing symbolic visualization tool, to assist with melody reduction and non-chord tone identification for computational musicology researchers and music theorists.

## Introduction

Automatic chord estimation (ACE) of symbolic music (i.e., the automatic labeling of polyphonic symbolic data) has become a popular task over the past decades in the field of computational musicology. Many chord recognition models have been built based on music theory rules, statistical approaches, or machine learning algorithms (Chen & Su, 2018; Masada & Bunescu, 2017; Temperley, 1997). However, a common obstacle across all three approaches is the appearance of non-chord tones in the score (Chen & Su, 2018; Masada & Bunescu, 2017; Radicioni & Esposito, 2010). Therefore, this paper focuses on building an automatic non-chord tone recognition model for a given melody in the common practice classical style.

In Western classical music theory, non-chord tones (NCTs) are categorized in many ways based on a melody note's metric position, and intervallic distance and direction from a prior and subsequent note (i.e., the intervals of arrival and departure) (Kostka & Payne, 2013, p.184). Each category of NCT is given a label based on their function (i.e., passing tone, neighbor tone, suspension, etc). To predict, then, whether a note is a chord tone (CT) in a melody and given no harmonic information, the important features are a note's arriving and departure intervals, metric position, and duration.

In this project, we propose a logistic regression model to distinguish NCTs and CTs using the Theme And Variation Encodings with Roman Numerals (TAVERN) corpus (Devaney et al., 2015). This model is then used to perform melodic reduction via non-chord tone identification for computational musicology researchers and music theorists.

### Background

With the prevalence of machine-learning approaches in the field of music information retrieval (MIR), several studies on ACE in symbolic music have indicated the presence of NCTs in a score add a significant amount of "noise" to a model, thus reducing accuracy. Prior studies have demonstrated relatively high prediction accuracy in datasets in strict chorale style (e.g., Bach) (Masada & Bunescu, 2017) which are known to contain mostly CTs. However, on non-homorhythmic and especially virtuosic styles, such as the TAVERN (Devaney et al., 2015) and Beethoven Piano Sonata (Chen & Su, 2018) corpora, model performances worsen. We believe NCT identification could be an essential task in the data processing stage for ACE in order to reduce the noisiness in a dataset.

A few researchers have attempted to either isolate NCTs directly--such as the *dissonant filter* tool in Verovio Humdrum Viewer (VHV) for identifying ornamental tones of different types (Sapp, 2017)—or else included NCTs as a feature in modeling harmony. For example, Temperley and Sleator used a rule-based system to identify ornamental dissonances in a model for predicting harmony labels (1999). However, both of these studies consider elements from a full polyphonic score in order to identify NCTs.

Recent research involving CT and NCT recognition has been more centered around statistical and machine learning models relying on parameters such as melodic features, segmentation, and vertical sonority (Condit-Schultz et al., 2018; Arthur 2016). Arthur (2016) created a backward stepwise regression model using duration, metric levels, approach and departure intervals only from melody to predict NCTs with the themes portion of the TAVERN corpus. The current project follows Arthur's research by exploring model performance using the whole TAVERN dataset. We additionally provide a visual implementation of the model to recognize NCTs in a melody more intuitively.

## Methods

The project methodology can be broken into two stages: data collection and pre-processing, and model fitting.

### Data Collection and Pre-processing

The data collection and pre-processing part contains two steps. The first is to extract musical features from the original score, to construct a data frame from the selected dataset, and the second step is to generate ground truth for each note based on its underlying labeled harmony in the corpus. The ground truth is simply whether or not a given melody note is a member of the underlying harmony or not (i.e., CT/NCT).

For this project we relied exclusively on the TAVERN corpus (Devaney et al., 2015),[2] which contains 27 sets of themes and variations (T&V) for piano by Mozart and Beethoven with 281 themes and variations in total. Each musical phrase is encoded as a separate *.krn* file containing the full score alongside Roman numeral and Functional interpretations.

We 'extracted' the melody using the Humdrum toolkit (Huron, 1995), which is a command-line tool that performs various tasks for symbolic music such as converting musical representations (e.g., pitch class to scale degree), extracting and calculating musical features (e.g., metric position, neighboring intervals, etc.), et cetera. Due to the complexity in tracing "the melody" for a given theme or variation, we extracted the upper-most voice and operationalized that as "melody" for the purposes of our study. With each melody note as a data point, a data frame was then constructed by extracting each note's scale degree under the current key, metric position, duration, arriving and departure intervals, along with the corresponding Roman numeral (RN) and the current key. The ground truth of each note—whether it is a chord tone or not—was then labeled via a NCT identification algorithm given its scale degree, associated RN, and the key. All information was then concatenated into one data frame in R for statistical evaluation.

The total number of eligible data points in the TAVERN dataset is 45, 299, with approximately 72% CTs and 28% NCTs. The variations subset of TAVERN has a higher percentage of NCTs compared with only 19% NCTs in the themes (Arthur, 2016). This is likely due to the stylistic tendency for variations to exhibit extreme ornamentation and elaboration techniques; these decorations are most commonly passing tones, neighboring tones, or appoggiaturas (Giraud et al., 2013).

### Model Fitting

We proposed a logistic regression model because the dependent variable—CT or NCT—is binary. Since Arthur's NCT model was only trained on the themes from the TAVERN dataset (Arthur, 2016), we wished to explore a model's performance trained on the whole dataset. As the distribution of CTs / NCTs in the dataset is skewed, the initial goal of the model was to demonstrate improvement over a baseline model that simply identifies every single note as a CT (i.e., accuracy is just the percentage of the CTs in the test set, or ~72%).

The independent variables in our model are: duration (linear), on-beat/off-beat (binary), arriving interval (coded as a step or a leap), and departure interval (step/leap). After splitting the TAVERN dataset into an 80/20% train and test set, we built a logistic regression model. Considering only the main effects, the accuracy on the test set is 75.33% with AUC = 0.78, which is higher than the baseline model's 70.30% accuracy (percentage of CTs in the test set). The AUC score in this case reflects the model's ability to distinguish between CT and NCT, where AUC = 1 represents perfect classification accuracy, while AUC = 0.5 implies zero discrimination ability. Since the testing accuracy will presumably increase when considering interactions between independent variables, we then applied a forward stepwise logistic regression until achieving the best AIC score, to find the variables and interactions that significantly impact the model's performance. The result of the model (referred as Model 1) with the best AIC (AIC = 34871.22) is shown in Table 1:

*Table 1: Significant Factors in Model 1.*

| Main Effects | Interactions |
|---|---|
| Departure Interval(DI) | AI:Dur |
| Arriving Interval(AI) | DI:AI |
| Duration(Dur) | AI:Beat |
| On/off-beat(Beat) | DI:Beat |
| | DI:Dur |
| | DI:Beat |
| | DI:AI:Beat |

The AIC of Model 1 is 34871.22, with an accuracy of 75.34% and an AUC of 0.79. Our next goal was to attempt to achieve an equal or higher testing accuracy using a simpler model. However, testing accuracy failed to improve as expected. A likely reason is the substantial noisiness in the variations as described earlier. Accordingly, we attempted to replicate Arthur's (2016) approach of training with only the themes (but testing on the full dataset). The logic here was that "rules" are more likely to be followed in simple themes but difficult to learn from the full variations. By applying the forward stepwise logistic regression, the best model (referred as Model 2) parameters are shown in Table 2:

*Table 2: Significant Factors in Model 2.*

| Main Effects | Interactions |
|---|---|
| Departure Interval(DI) | AI:Beat |
| Arriving Interval(AI) | DI:Beat |
| Duration(Dur) | DI:Dur |
| On/off-beat(Beat) | |

The AIC of Model 2 is 1202.98, and its accuracy on the test set is 75.39% with an AUC of 0.78. Model 2 successfully replicates Arthur's findings (Arthur, 2016). With themes as the training data, though the accuracy and AUC did not significantly improve, the model is much simpler with a substantially lower AIC. In other words, we achieved the same accuracy with a simpler model. As a result, we further investigated Model 2 in the following subsections.

*Exploring Other Factors*

Besides duration, on-beat/off-beat, arriving and departure intervals, Arthur's model also included whether a note is a boundary note, referring to whether a note is a starting or an ending note of a piece, as another factor. Accordingly, we not only introduced a new variable, boundary note (binary), but also included whether a note was arriving by a rest (binary) or departing to a rest (binary) along with the previous factors. After repeating training on the themes dataset using forward stepwise logistic regression, the best model was exactly the same as Model 2, allowing us to conclude that these new factors do not contribute to the performance of the model.

**Results**

This subsection evaluates the performance and generalizability of Model 2 by performing cross-validation and testing on other classical music datasets. Cross-validation is applied to test the model's ability to classify CTs/NCTs in new (unseen) data, and addresses problems like overfitting or selection bias, and gives insights on the model's overall generalizability. As a first litmus test, we wanted to test Model 2 on random subsets of the themes portion of the dataset itself. The themes portion of the dataset has 2,039 data points with 81% CTs and 19% NCTs. Our logic was that given the simplicity and uniformity in the themes compared with the variations, the model ought to perform better. Using an 80%/20% split training and testing, we performed a 10-fold cross-validation on the themes data, the results of this test of Model 2 (using parameters shown in Table 2) is shown in Table 3.

*Table 3: Cross-validation of Model 2 on Themes Only.*

| **Percentage of chord tones in the training set**: 83.19% |
|---|
| **Cross-validation**: |
| Accuracy: 84.87% |
| Precision: 0.8756 |
| Recall: 0.9508 |
| f1: 0.9116 |
| AUC: 0.8747 |

The results indicate that the model makes sufficient predictions on the themes dataset itself. Of course, it is still biased due to the extreme proportion of CTs. To examine Model 2's generalizability, we applied the model on other datasets as well. Previously we tested the model on only a part of the TAVERN dataset (i.e., a fixed 20% test set). Since the melodic complexity can vary significantly across different variations, we use cross-validation to re-test the model on the complete TAVERN dataset by randomly selecting 2,400 data points over 50 separate trials and take the average over all trials (Table 4). In addition to TAVERN, we test Model 2 on another dataset with harmonic annotations of Joseph Haydn's "Sun Quartets" dataset in *kern* format (López, 2017), which contains 6 string quartets (12,616 data points). The Model 2 results tested on the two datasets (full TAVERN and Sun Quartets) are displayed in Table 4. The prediction accuracy is higher than the baseline model in both cases, and Model 2 is capable of distinguishing CTs and NCTs in the melodies from the two datasets.

*Table 4: The Baseline accuracy, predicting accuracy, and AUC on different datasets.*

| Dataset | TAVERN | Haydn |
|---|---|---|
| **Baseline** | 71.2% | 66.0% |
| **Model 2** | 76.4% | 70.6% |
| **AUC** | 0.7766 | 0.6852 |

So far, we investigated and evaluated the best model trained with the themes dataset (Model 2), using the full set of parameters as shown in Table 2. In the next subsection, we introduce a visualization tool for melodic reduction using our model.

*Visualization in Verovio Humdrum Viewer*
This section presents an application of the model to a score visualization program, using color to allow easy identification of notes predicted as NCTs in a melody. We use the Verovio Humdrum Viewer (VHV) as our visualization platform, which is an online symbolic music editor and interactive notation rendering interface for *.krn* and *.mei* files (Sapp, 2017).

The data processing of a new score is similar to the procedure in Section *Data Collection and Pre-processing*. After feeding the data from the new score into our model, users can modify a probability threshold after which the model generates prediction probabilities for each note. The default threshold is 0.5, which takes probability larger than 0.5 as CTs and smaller than 0.5 as NCTs. Next, the program creates another spine **color* in the original score to distinguish CTs and NCTs: CTs are labeled as black where NCTs are labeled as pink by matching index. An example is shown in Figure 1. Note that RNs are not used in the model and are only shown as a visualization aid.

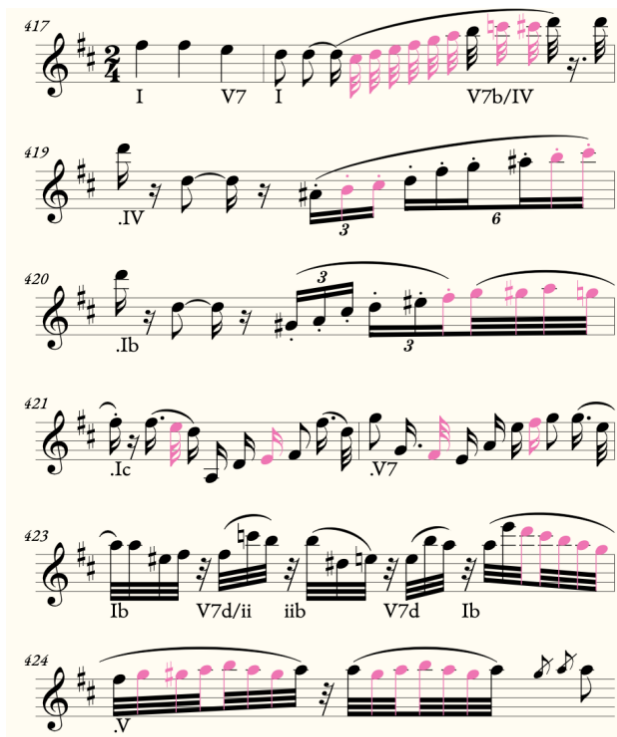

*Figure 1: Visualization of a phrase in the TAVERN dataset (with harmony labeled), where black notes are chord tones and pink notes are non-chord tones.*

The visual implementation of the model in VHV provides an intuitive way to recognize NCTs in a melody or to evaluate our model's performance. The visualization would assist computational musicologists and music theorists with music analysis.

## Conclusion
This paper proposed a logistic regression model to predict non-chord tones (NCTs) from a single melodic line from a classical score. The best model, trained on the themes of the TAVERN dataset, used several melodic features as NCT predictors (i.e., duration, beat, arriving and departure intervals). The model can successfully distinguish CTs and NCTs across multiple datasets with an average accuracy of about 73.5%. Finally, we introduced a visualization tool using the model, which performs melodic reduction on a monophonic symbolic score for computational musicology researchers and music theorists.

## Discussion
Future improvements will include simplified user experience of the visualization program, building a web page or a pipeline tool, and/or implementing the function as a permanent feature in Verovio Humdrum Viewer. Our model does not guarantee the best performance for several reasons. First, since the ratio of NCT to CT are usually unbalanced in classical music this leads to a higher possibility for a model to predict a note as CT. Second, our ground truth is based on human-annotated Roman numerals, which, despite being 'expert' is far from an objective process. For instance, Koops et al. (2020) indicated some annotators may prefer to annotate fewer chord changes to emphasize the longer-range structure of the piece, while others may analyze harmonies in more detail. Thus, the ground truth of a dataset—whether a note is an CT/NCT—impacts the information a model is trained on, which therefore will further affect the performance of a model. Finally, our operationalization of "melody" was oversimplified, and longer-range contextual features (beyond n=2) were not considered.

Other applications of the model would include expanding to apply it to other genres. Our eventual goal is to use the model as a melody reduction process to potentially improve ACE accuracy in symbolic music. We are currently working on implementing the model as a preprocessing stage for ACE in polyphonic symbolic music using a deep learning method. The study focuses on classical string quartets, applying the model to each independent voice in order to perform a full score reduction. Our preliminary results indicate a modest improvement in recognition accuracy over a baseline model (feedforward neural network) after applying score reduction. We look forward to presenting the work in the near future.

## End Notes
[1] Project code at
https://github.com/TianxueHu/Auto_NCT
[2] The TAVERN Corpus:
https://github.com/musicnerd/TAVERN